\documentclass[twocolumn,floatfix,showpacs,prl,10pt, superscriptaddress]{revtex4-1}
\usepackage{graphicx}

\usepackage{amsmath}
\usepackage{amssymb}
\usepackage{bm}
\usepackage{hyperref}
\usepackage{multirow}
\usepackage{color}
\usepackage{bbold}
\usepackage[usenames,dvipsnames]{xcolor}

\usepackage{amsthm}
\usepackage{amsbsy}
\usepackage{natbib}

\newcommand{\pd}{{\phantom\dag}}

\begin{document}

\title{Aperiodic Weak Topological Superconductors}

\author{I. C. Fulga}
\affiliation{Department of Condensed Matter Physics, Weizmann Institute of Science, Rehovot 76100, Israel}
\author{D. I. Pikulin}
\affiliation{Department of Physics and Astronomy and Quantum Matter Institute, University of British Columbia, Vancouver, BC, Canada V6T 1Z1}
\author{T. A. Loring}
\affiliation{Department of Mathematics and Statistics, University of New Mexico, Albuquerque, NM 87131, USA}

\date{\today}
\begin{abstract}
Weak topological phases are usually described in terms of protection by the lattice translation symmetry. 
Their characterization explicitly relies on periodicity since weak invariants are expressed in terms of the momentum-space torus.
We prove the compatibility of weak topological superconductors with aperiodic systems, such as quasicrystals.
We go beyond usual descriptions of weak topological phases and introduce a novel, real-space formulation of the weak invariant, based on the Clifford pseudospectrum.
A non-trivial value of this index implies a non-trivial bulk phase, which is robust against disorder and hosts localized zero-energy modes at the edge.
Our recipe for determining the weak invariant is directly applicable to any finite-sized system, including disordered lattice models.  This direct method enables a quantitative analysis of the level
of disorder the topological protection can withstand.
\end{abstract}
\maketitle

\emph{Introduction} ---
One of the hallmarks of a topological phase is the presence of quantized macroscopic observables which are insensitive to perturbations or random modulations of the local microscopic environment \cite{Hasan2010, Qi2011}. This remarkable feature can be understood by expressing the quantized observables as topological invariants of an underlying microscopic theory of the bulk system, such that they are unchanged by arbitrary local deformations which do not close the bulk gap. The most well-known example is the transverse conductivity of the quantum Hall effect \cite{Klitzing1980}: owing to the topological protection \cite{Thouless1982}, it shows a level of quantization which is used to define the metrological standard of resistance \cite{Klitzing1985}.

In many cases however, topological protection requires restricting the space of allowed perturbations, leading to so-called symmetry protected topological phases \cite{Kitaev2009, Schnyder2009, Wen2014}. The latter can also be described in terms of topological invariants, similar to the quantum Hall effect, but only under the condition that all perturbations respect a certain set of symmetries. Imposing time-reversal, particle-hole, or chiral symmetries leads to non-trivial phases called strong topological insulators (TI), while the requirement of translation or point group symmetries of the lattice leads to weak and crystalline topological phases (WTI, TCI), respectively \cite{Fu2007a, Moore2007, Roy2009, Fu2011, Hsieh2012}.
This terminology, weak versus strong, refers to the accuracy with which the protecting symmetries can be enforced in a realistic setting. In a time-reversal invariant strong topological insulator, contamination by magnetic impurities can be controlled in experiment. In a weak topological phase however, translation symmetry will always be broken by impurities and random displacements of atoms in a crystal.

Nevertheless, even when disorder breaks the lattice symmetry, a WTI or TCI can still be robust \cite{Nomura2008, Ringel2012, Mong2012, Fu2012, Kobayashi2013, Fulga2014, Baireuther2014, Sbierski2014, Diez2015, Morimoto2015, Milsted2015, Song2015, Fulga2015}. It was found that topological invariants can still be defined and the boundary can avoid localization when the system remains symmetric on average, that is to say the full ensemble of disorder configurations is invariant under the symmetry \cite{Fulga2014}. Due to self-averaging, the topological invariant approaches its quantized value as the system becomes larger and explores more of the ensemble of disorder configurations \cite{Song2015}. In effect, even though a disordered WTI locally breaks translation symmetry, the latter is restored for the purpose of finding the topological invariant on macroscopic length scales.

So far, all studies of disordered WTI used as a starting point a \emph{clean system} discretized on a lattice -- \emph{i.e.} a periodic arrangement of sites. From this starting point, it is shown that the phase is robust to adding random perturbations which break the lattice translation symmetry. 
Such an approach is convenient since most expressions for weak invariants explicitly rely on periodicity. Indeed, homotopy theory \cite{Kennedy2015} and K-theory \cite{Kitaev2009, Shiozaki2014} describe WTI phases in terms of invariants over the momentum-space torus.
But there are cases in which this separation -- lattice Hamiltonian plus random perturbation -- cannot be made, since there is no lattice: these are aperiodic systems, such as quasicrystals \cite{Baake2013}.

We consider a two-dimensional (2d) topological superconductor on an aperiodic tiling, and prove by construction that it can host a weak topological phase, similar to lattice systems. Our work provides the first example of a new class of non-trivial Hamiltonians which cannot be directly characterized by conventional weak indices, owing to the absence of momentum-space even in the clean limit. One can presumably approximate the Hamiltonian by periodic Hamiltonians with defects and use
the momentum torus method for defining a weak topological phase given a 
mobility gap, as was done for strong topological phases in \cite{BandresPhotonicQQ}. Due to the defects induced in the periodic approximation, a subsequent quantitative analysis of the degree
of topological protection would probably be imprecise.

Instead, we introduce a novel, real-space formulation of the weak invariant, based on the Clifford pseudospectrum \cite{Loring2015}, and confirm its validity by using scattering theory \cite{Fulga2011, Fulga2012}. We prove that a non-trivial value of this weak index implies the presence of a gapped \emph{bulk phase}, which is topologically non-trivial in a weak sense and robust against disorder. This means that beyond introducing aperiodic weak topological superconductors, our work provides a concrete and quantitative recipe for determining weak topological invariants without using momentum-space, which may be applied to any system, including lattice Hamiltonians with or without disorder. 
As such, our method is a practical and timely alternative for determining weak topological effects, which have been the focus of an increasing number of material proposals and experimental studies \cite{Yan2012, Rasche2013, Tang2014, Yang2014, Pauly2015, Autes2015}.

\emph{Quasicrystalline topological superconductor} --- 
Models of two-dimensional topological superconductors with broken time-reversal symmetry usually describe spinless fermions on a lattice in the presence of $p$-wave odd-momentum pairing, $\Delta({\bf p}) = -\Delta(-{\bf p})$ \cite{Asahi2012, Seroussi2014, Buehler2014, Diez2014}. The minimal Bogoliubov de Gennes (BdG) Hamiltonian takes the form
\begin{equation}\label{eq:h_pwave}
 H_{\rm BdG}({\bf p}) = \left( \frac{p_x^2+p_y^2}{2m} - \mu \right) \sigma_z + \Delta p_x \sigma_x + \Delta p_y \sigma_y,
\end{equation}
where $p_{x,y}$ are the two momenta, $m$ is the effective mass, $\mu$ the chemical potential, and $\Delta$ the strength of the $p$-wave pairing. The Pauli matrices $\sigma_i$ parametrize the particle-hole degree of freedom. The Hamiltonian \eqref{eq:h_pwave} obeys a particle-hole symmetry (PHS) of the form
$
 H^\pd_{\rm BdG}({\bf p}) = - \sigma_x H^*_{\rm BdG}(-{\bf p}) \sigma_x,
$
with an anti-unitary PHS operator ${\cal P} = \sigma_x {\cal K}$, where ${\cal K}$ is complex conjugation. Since ${\cal P}^2=+1$, $H_{\rm BdG}$ belongs to class D in the Altland-Zirnbauer classification \cite{Altland1997}.

\begin{figure}[tb]
\includegraphics[width=0.95\columnwidth]{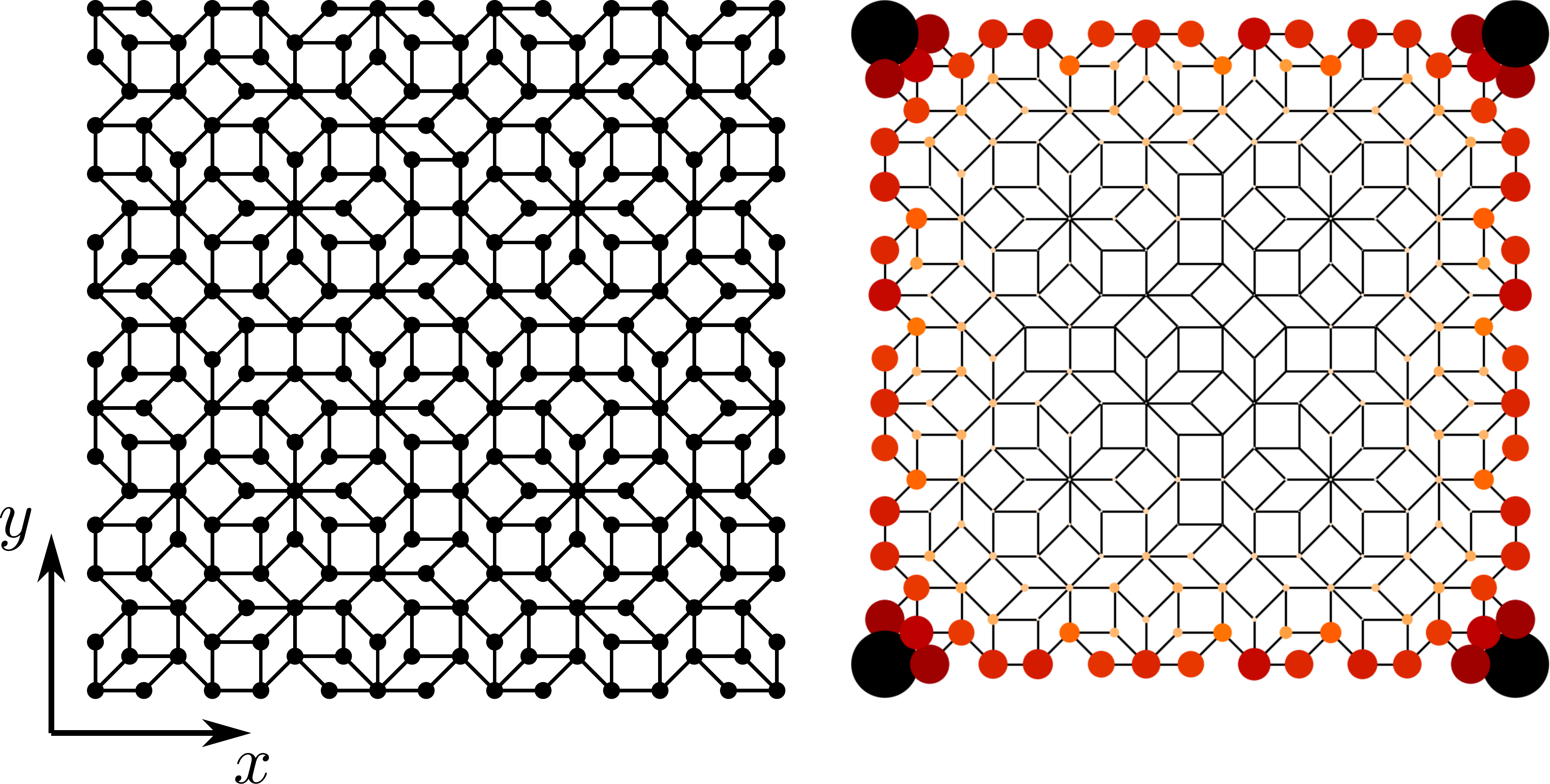}
\caption{Left: Patch of the Ammann-Beenker tiling in the $x$ - $y$ plane, obtained by repeated subdivision of a square \cite{apptiling}. A tight binding model $H_{QC}$ is defined by associating an on-site Hamiltonian and a hopping matrix to each vertex and link in the patch, respectively. Right: Total wavefunction amplitude of $H_{QC}$, corresponding to states with energies $|E| < 0.2$, for $t=\Delta=1$ and $\mu=2$. Circles of larger area and darker color correspond to larger amplitudes.\label{fig:ammann}}
\end{figure}

Our aim is to obtain a similar model on a quasicrystal, such as the 2d Ammann-Beenker tiling shown in Fig.~\ref{fig:ammann}. To this end, we construct a real-space tight binding model by associating a Hamiltonian term to each site of the tiling, and a hopping matrix to each link between neighboring sites. The Hamiltonian on-site element corresponding to site $j$ reads
\begin{equation}\label{eq:h_ons}
 H_j = -\mu \sigma_z
\end{equation}
and the hopping matrix between neighboring sites $j$ and $k$ is given by\begin{equation}\label{eq:h_hop}
 H_{jk} = -t \sigma_z - \frac{i}{2}\Delta\sigma_x \cos(\alpha_{jk}) -\frac{i}{2}\Delta\sigma_y \sin(\alpha_{jk}).
\end{equation}

In Eqs.~\eqref{eq:h_ons} and \eqref{eq:h_hop} $\mu$ and $\Delta$ have the same meaning as in \eqref{eq:h_pwave}, with $t$ the hopping strength, and $\alpha_{jk}$ the angle of the bond between site $j$ and site $k$, measured with respect to the horizontal direction.
The resulting tight binding Hamiltonian takes the form
\begin{equation}\label{eq:hqc}
 H_{\rm QC} = \sum_j {\bf c}^\dag_j H^\pd_j {\bf c}^\pd_j + \sum_{\langle j,k\rangle} {\bf c}^\dag_j H^\pd_{jk} {\bf c}^\pd_k,
\end{equation}
where the second term is summed over neighboring sites, with ${\bf c}^\dag_j = (c^\dag_j, c^\pd_j)$, and $c^\pd_j$ the fermionic annihilation operator at site $j$. $H_{\rm QC}$ still obeys particle-hole symmetry, which in real-space reads
\begin{equation}\label{eq:qc_phs}
 \Sigma_x H^\pd_{\rm QC} \Sigma_x = - H_{\rm QC}^*,
\end{equation}
where the block-diagonal matrix $\Sigma_x = \sigma_x \oplus \sigma_x \oplus \cdots \oplus \sigma_x$.

If the 2d Hamiltonian $H_{\rm QC}$ would describe a system on an infinite square lattice, its momentum-space form would be given by \eqref{eq:h_pwave} up to a rescaling of the chemical potential. Even though this relation no longer holds in an aperiodic tiling, $H_{\rm QC}$ still belongs to symmetry class D, as a consequence of the constraint \eqref{eq:qc_phs}. Therefore, it allows for a topological classification in terms of the Chern number, in which topologically non-trivial phases are characterized by chiral propagating Majorana edge modes. We study the system numerically using the Kwant code \cite{Groth2014, sourcecode}, finding a gapped bulk and gapless boundary states at the Fermi level, $E=0$, for $t=\Delta=1$ and $\mu=2$ (see Fig.~\ref{fig:ammann}, right panel).

We test the nature of the edge states by using scattering theory. Attaching two infinite, translationally invariant leads to the left- and right-most sites of the patch in Fig.~\ref{fig:ammann} allows to compute the Fermi level scattering matrix \cite{appsmatrix},\begin{equation}\label{eq:smatrix}
 S = \begin{pmatrix}
      r & t \\
      t' & r'
     \end{pmatrix},
\end{equation}
where the $t^{(\prime)}$ and $r^{(\prime)}$ blocks contain the transmission and reflection amplitudes of the lead modes, respectively. From the scattering matrix, we obtain the thermal conductance in the low-temperature, linear response regime: $G = G_0{\rm Tr}\,t^\dag t$, with $G_0 = \pi^2 k_B^2 T_0 / 6h$ the quantum of thermal conductance.

For the parameters of Fig.~\ref{fig:ammann}, the thermal conductance is quantized, $G/G_0=1$, a hallmark of a topological superconductor with Chern number $|C|=1$. 
To confirm the topological origin of this edge mode, we compute the pseudospectrum $\mathbb{Z}$-index introduced in Ref.~\cite{Loring2015}. Performing a change of basis, $\widetilde{H}_{QC} = \Omega H_{QC} \Omega^\dag$, with $\Omega = A \oplus A \oplus \cdots \oplus A$ and
\begin{equation}\label{eq:basischange}
 A= \sqrt{\frac{1}{2}} \begin{pmatrix}
    1 & 1 \\
    -i & i
   \end{pmatrix},
\end{equation}
leads to an imaginary Hamiltonian, $\widetilde{H}^\pd_{QC} = - \widetilde{H}_{QC}^*$. The strong pseudospectrum invariant can then be obtained as
\begin{equation}\label{eq:strong_inv_ps}
 C_{ps} = \frac{1}{2}{\rm Sig} \begin{pmatrix}
                     X & Y - i \widetilde{H}_{QC} \\
                     Y + i \widetilde{H}_{QC} & -X
                    \end{pmatrix}.
\end{equation}
Here, $X$ and $Y$ are the position operators associated to the sites of the tiling (see Fig.~\ref{fig:ammann}) and ${\rm Sig}$ stands for matrix signature, \emph{i.e.} the number of positive eigenvalues minus the number of negative eigenvalues.    Choosing site coordinates which span $-1/2\leq x\leq 1/2$ and $-1/2\leq y \leq 1/2$ gives a pseudospectrum invariant $C_{ps}=-1$, consistent with the quantized value of the thermal conductance. The index $C_{ps}$ is well defined since the matrix of Eq.~\eqref{eq:strong_inv_ps} does not have eigenvalues close to zero for our choice of parameters. We provide a more detailed analysis of this invariant in the Supplemental Material.

\emph{Weak topological superconductor without a lattice} --- 
The presence of a Chern insulating phase in the quasicrystalline system $H_{\rm QC}$ is not surprising. The tenfold classification of topological insulators and superconductors shows that such a phase can appear in 2d systems both with and without the PHS constraint \eqref{eq:qc_phs}, without requiring any spatial symmetries. In fact, aperiodic Chern insulators without PHS (symmetry class A) were studied in Refs.~\cite{Tran2015, BandresPhotonicQQ}.

Weak topological phases on the other hand are typically described as being protected by the translational symmetry of the lattice. A WTI phase in a 2d class D system can be thought of as a set of parallel, weakly coupled Kitaev chains \cite{Kitaev2001}. Each Kitaev chain in the array is a strong 1d topological superconductor, hosting an unpaired Majorana zero mode at each end irrespective of spatial symmetries. In the 2d coupled system however, topological protection is ensured by translation symmetry in the direction perpendicular to the chains, or by average translation symmetry for a disordered lattice.

\begin{figure}[tb]
 \includegraphics[width=\columnwidth]{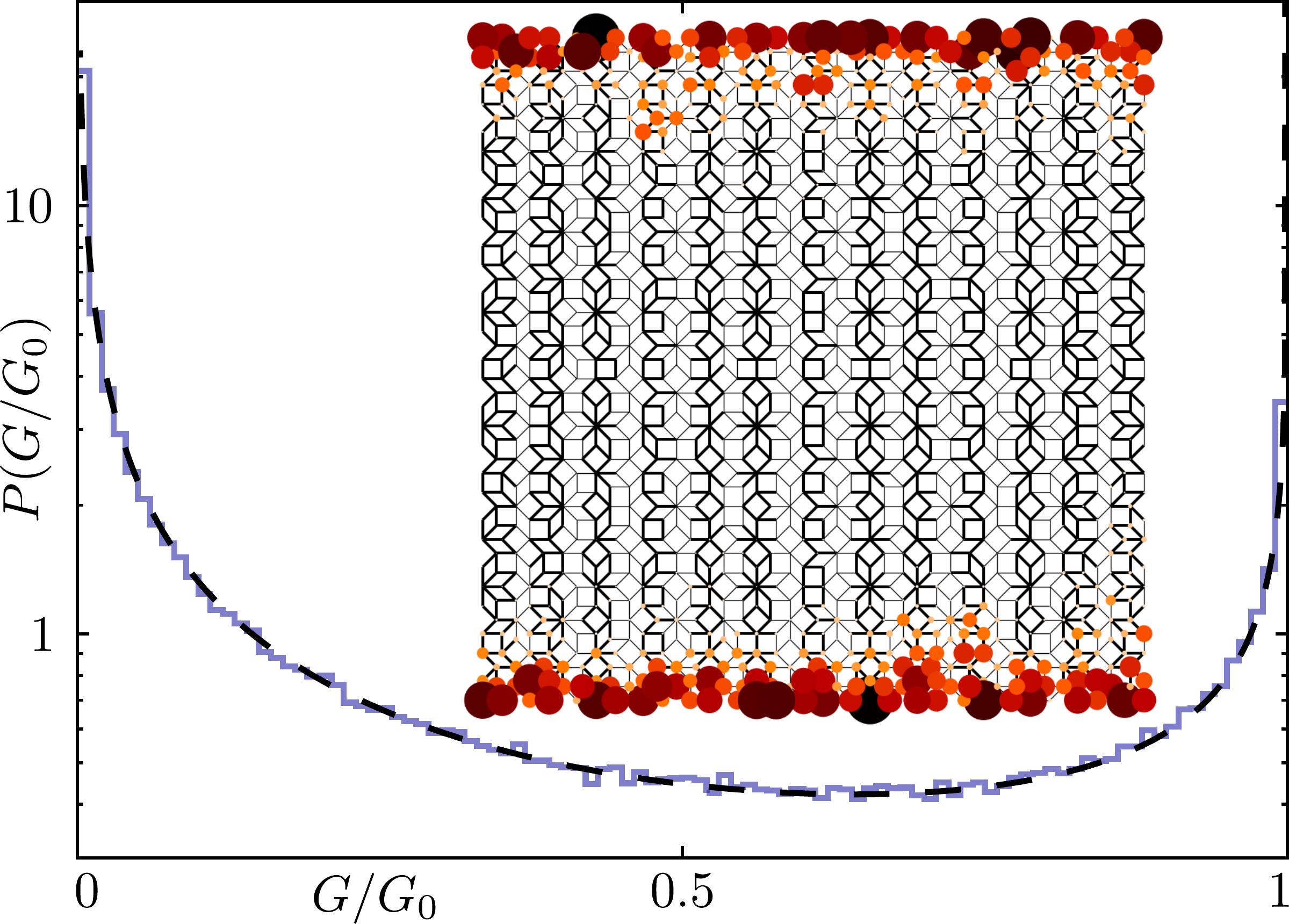}
 \caption{Log-linear plot of the thermal conductance distribution of a single edge in the WTI phase. The histogram (solid line) is computed numerically from $2\times10^5$ disorder realizations, using $t=\Delta=U=1$ and $\mu=1.9$, for a quasicrystal patch composed of 13 weakly coupled wires. The dashed line shows the analytic result of Eq.~\eqref{eq:bimodal_distro}, using $l/L=0.18$. Inset: total amplitude of wavefunctions with energies $|E|<0.1$ for a single disorder realization. Circles of larger area and darker color correspond to larger amplitudes, while thicker hoppings show the positions of Kitaev chains in the array.\label{fig:qc_wti}}
\end{figure}

We seek to produce such a phase in the quasicrystalline tight binding model $H_{\rm QC}$. In order to convert the aperiodic tiling into an array of Kitaev chains, we selectively reduce hopping amplitudes in regions of the quasicrystal, multiplying some of the hopping matrices \eqref{eq:h_hop}, $H_{jk}$, by a global factor of $0.2$. We set $t=\Delta=1$, $\mu=1.9$, and form a total of $13$ weakly coupled quasi-1d strips, as shown in Fig.~\ref{fig:qc_wti}. While Fig.~\ref{fig:ammann} showed a strong topological phase in which all boundaries host mid-gap states and $C_{ps}=-1$, here only the top and bottom boundaries are gapless and $C_{ps}=0$, characteristic of weak topological phases.

For these Hamiltonian parameters, each quasi-1d strip becomes a non-trivial Kitaev chain, and its Majorana zero modes couple to those of neighboring chains, leading to the formation of two Kitaev edges \cite{Diez2014}. 
In a lattice system, the presence of exact translational symmetry implies equal coupling between all adjacent Majorana end states, leading to a pair of decoupled, counter-propagating modes at each edge. The total thermal conductance of the system then becomes $G/G_0=2$, having a quantized, unit contribution from the conducting mode at each of the two edges. As before, we attach leads to the leftmost and rightmost sites and compute the thermal conductance, finding $G/G_0=1.94$. The deviation from a quantized value means that counter-propagating modes of the two Kitaev edges are coupled. Nevertheless, it was shown that an edge can still avoid localization provided the couplings between adjacent Majoranas are independent and statistically equivalent (they all have the same probability distribution). In this case, each of the two Kitaev edges is pinned to a 1d topological phase transition \cite{Brouwer2000, Motrunich2001, Brouwer2003, Gruzberg2005}, being characterized by a bimodal conductance distribution \cite{Diez2014}:
\begin{equation}\label{eq:bimodal_distro}
\begin{split}
P(G/G_0) &= \sqrt{\frac{l}{2\pi L}} (G/G_0)^{-1} (1 - G/G_0)^{-1/2} \\
 & \times \exp \left( -\frac{l}{2L} {\rm arccosh}^2 \sqrt{\frac{1}{G/G_0}}\, \right),
\end{split}
\end{equation}
peaked at $G/G_0=0$ and $1$, where $l$ is the mean free path, and $L$ is the length of the edge.

We test the localization properties of the aperiodic Kitaev edge by adding on-site as well as hopping disorder terms to $H_{\rm QC}$. We replace $\mu \to \mu+\delta_\mu$, with $\delta_\mu$ drawn randomly for each site from the uniform distribution $[-U,U]$, where $U$ is the strength of disorder. Additionally, we add randomness in the coupling between the Kitaev chains, multiplying each of the hopping matrices connecting them by a random factor drawn uniformly from the interval $[0.1,0.3]$. By attaching leads to the left, right, as well as bottom sites, we determine the thermal conductance contribution of only the top edge \cite{appsmatrix}.
The numerical results shown in Fig.~\ref{fig:qc_wti} closely follow the analytical prediction of Eq.~\eqref{eq:bimodal_distro}, implying a lack of localization.

Remarkably, the system behaves as a WTI even though it does not have a lattice. On the level of the aperiodic tiling, one cannot define an exact or average translation symmetry, owing to the absence of Bravais vectors even in the clean limit. As such, conventional momentum-based expressions for the weak invariants cannot be directly applied, short of restoring periodicity. The latter can be achieved, for instance, by imposing twisted boundary conditions on a finite tiling with suitably chosen terminations, thereby forming a torus \cite{Essin2007}. Another possibility is based on unpublished work by Kitaev \cite{Kitaevlecture}, which shows that for sufficiently large systems there exists a coarse-graining procedure transforming the Hamiltonian into a Dirac-like operator with a mass term. While generic enough to apply to systems in any dimension and symmetry class, Kitaev's approach does not offer a recipe for performing the coarse-graining, such that it can be applied to specific systems.

We seek a practical yet quantitative method for determining weak invariants in the absence of momentum-space, such that it can be applied in an aperiodic setting.
Our real-space formulation is based, as before, on the Clifford pseudospectrum, as well as on the observation that weak indices can be expressed as strong invariants of lower dimensional systems. Therefore, we ignore the $x$ coordinate and use the 1d form of the pseudospectrum invariant in class D \cite{Loring2015}:
\begin{equation}\label{eq:weak_inv_ps}
 Q_y = {\rm sign}\,\det \left( Y + i\widetilde{H}_{\rm QC} \right).
\end{equation}

The matrix in Eq.~\eqref{eq:weak_inv_ps} is real since the Hamiltonian $\widetilde{H}_{\rm QC}$ is imaginary after the basis change \eqref{eq:basischange}. For an array composed of an odd number of Kitaev chains like the one in Fig.~\ref{fig:qc_wti}, we find a non-trivial value of the $\mathbb{Z}_2$ index: $Q_y=-1$. As an independent confirmation of the validity of this expression, we compute the weak invariant also using scattering theory. By attaching leads only to the top and bottom sites of the system, such that they contact the two Kitaev edges, the weak index can be obtained from the reflection block of the scattering matrix \eqref{eq:smatrix} as $\nu_y = {\rm sign}\,\det\, r$ \cite{Akhmerov2011, Fulga2012, nopbchere}. The value of the scattering matrix invariant is consistent with that obtained from the pseudospectrum, $\nu_y=Q_y=-1$.

A nontrivial value of $Q_y$ forces topologically protected approximate zero modes to exist near the two edges of the system corresponding to positions $y=\pm 0.5$. This follows from the two statements below. First, we can prove that for the clean Hamiltonian with a non-trivial weak index in the bulk \eqref{eq:weak_inv_ps}, there is a position near the edge of the sample as a function of $y$ where the gap closes and there is a zero-energy state localized in $y$. Second, if on-site disorder is on average not larger than the gap from the Fermi energy to the bulk states, the zero energy state survives the effect of disorder, although the expected $y$ position may move in somewhat from the edge. These statements are made precise as Corollary~1 in the Supplemental Material. The proofs of these statements follow rather directly from the results in \cite{Loring2015} applied to the 1d system $(Y,H_{\rm QC})$.  

\emph{Discussion and future work} --- 
The gapless boundary states of weak topological insulators and topological crystalline insulators are typically thought of as a consequence of lattice symmetries. In all rigorous descriptions of such phases, either using K-theory or homotopy theory, the lattice periodicity is explicitly built in from the start: invariants are defined using the momentum-space torus. Our work goes beyond this paradigm, and instead considers quasicrystalline weak topological insulators, systems in which there is no lattice.

We have proven, by construction, that a weak phase can exist on an aperiodic tiling. Throughout its characterization we have purposely avoided concepts relating to periodicity, such as Bloch states or the Brillouin zone. Instead, we have introduced a real-space expression for the weak index, based on the Clifford pseudospectrum. A non-trivial value of the invariant implies a non-trivial bulk phase, which is topologically equivalent to a weak phase on a periodic lattice \cite{Diez2014} and hosts gapless boundary states. The robustness of these boundary states is certainly greater than implied by Corollary~1. For example,
we expect our weak invariant to be insensitive to perturbations away from the center in $y$ so each boundary state should be relatively immune to large perturbations at the opposite edge.

Our work provides the first example of a new class of Hamiltonians which host weak topological phases in the absence of a lattice. While strong topological invariants can be studied without using periodicity \cite{Kellendonk2015}, this is not the case for weak phases. Current descriptions of weak invariants in disordered systems assume momentum to be a good quantum number in the clean limit \cite{Fulga2014, Song2015}, so they cannot be directly applied. One available option is to artificially restore periodicity, by imposing twisted boundary conditions and forming a torus \cite{Essin2007}, or by defining a coarse-grained Hamiltonian equivalent to a Dirac-like operator \cite{Kitaevlecture}. In the absence of momentum-space however, pseudospectrum invariants are up to now the only viable and quantitative tool \cite{Loring2015}.

Furthermore, for a finite-sized system, our method proves to be computationally more efficient than previous Hamiltonian expressions for the weak invariant. Determining the momentum-space weak invariant requires imposing twisted boundary conditions to restore periodicity, and computing the Pfaffian of the Hamiltonian at two points: $k=0$ and $k=\pi$. In contrast, the pseudospectrum invariant of Eq.~\eqref{eq:weak_inv_ps} only requires computing the determinant once, and allows to determine the topological features of systems with complex boundary shapes. We hope this method will become a standard tool in analyzing the robustness of weak topological phases.

While our work is based on a specific tiling and symmetry class, a generic, rigorous treatment of aperiodic weak phases and their relation to strong phases provides an interesting direction for future research.
Beyond the classification of topological phases,
the classification of quasicrystals themselves is still ongoing; whether different tilings can host weak phases with properties not found in lattice systems is an open question. For instance, quasicrystals can exhibit rotation symmetries which are impossible in periodic systems, raising the question of whether these symmetries can lead to protected boundary states. If so, the resulting phases would only be realizable in aperiodic tilings.

\begin{acknowledgments}

We are grateful to A. R. Akhmerov, M. Diez, H. Schultz-Baldes and J. Tworzyd{\l}o for useful discussions. This work was supported by a grant from the Simons Foundation (208723 to Loring), NSERC, CIfAR and Max Planck - UBC Centre for Quantum Materials, the European Research
Council under the European Union’s Seventh Frame-
work Programme (FP7/2007-2013) / ERC Project MU-
NATOP, the US-Israel Binational Science Foundation
and the Minerva Foundation.
\end{acknowledgments}


\begin{thebibliography}{0}%
\makeatletter
\providecommand \@ifxundefined [1]{%
 \@ifx{#1\undefined}
}%
\providecommand \@ifnum [1]{%
 \ifnum #1\expandafter \@firstoftwo
 \else \expandafter \@secondoftwo
 \fi
}%
\providecommand \@ifx [1]{%
 \ifx #1\expandafter \@firstoftwo
 \else \expandafter \@secondoftwo
 \fi
}%
\providecommand \natexlab [1]{#1}%
\providecommand \enquote  [1]{``#1''}%
\providecommand \bibnamefont  [1]{#1}%
\providecommand \bibfnamefont [1]{#1}%
\providecommand \citenamefont [1]{#1}%
\providecommand \href@noop [0]{\@secondoftwo}%
\providecommand \href [0]{\begingroup \@sanitize@url \@href}%
\providecommand \@href[1]{\@@startlink{#1}\@@href}%
\providecommand \@@href[1]{\endgroup#1\@@endlink}%
\providecommand \@sanitize@url [0]{\catcode `\\12\catcode `\$12\catcode
  `\&12\catcode `\#12\catcode `\^12\catcode `\_12\catcode `\%12\relax}%
\providecommand \@@startlink[1]{}%
\providecommand \@@endlink[0]{}%
\providecommand \url  [0]{\begingroup\@sanitize@url \@url }%
\providecommand \@url [1]{\endgroup\@href {#1}{\urlprefix }}%
\providecommand \urlprefix  [0]{URL }%
\providecommand \Eprint [0]{\href }%
\providecommand \doibase [0]{http://dx.doi.org/}%
\providecommand \selectlanguage [0]{\@gobble}%
\providecommand \bibinfo  [0]{\@secondoftwo}%
\providecommand \bibfield  [0]{\@secondoftwo}%
\providecommand \translation [1]{[#1]}%
\providecommand \BibitemOpen [0]{}%
\providecommand \bibitemStop [0]{}%
\providecommand \bibitemNoStop [0]{.\EOS\space}%
\providecommand \EOS [0]{\spacefactor3000\relax}%
\providecommand \BibitemShut  [1]{\csname bibitem#1\endcsname}%
\let\auto@bib@innerbib\@empty
\end{thebibliography}%


%


\begin{thebibliography}{99}
 \bibitem{Hasan2010} M. Z. Hasan and C. L. Kane, Rev. Mod. Phys. {\bf 82}, 3045 (2010).
 
 \bibitem{Qi2011} X.-L. Qi and S.-C. Zhang, Rev. Mod. Phys. {\bf 83}, 1057 (2011).
 
 \bibitem{Klitzing1980} K. v. Klitzing, G. Dorda, and M. Pepper, Phys. Rev. Lett. {\bf 45}, 494 (1980).
 
 \bibitem{Thouless1982} D. J. Thouless, M. Kohmoto, M. P. Nightingale, and M. den Nijs, Phys. Rev. Lett. {\bf 49}, 405 (1982).

 \bibitem{Klitzing1985} K. v. Klitzing and G. Ebert, Metrologia {\bf 21}, 1118 (1985).
 
 \bibitem{Kitaev2009} A. Y. Kitaev, AIP Conf. Proc. {\bf 1134}, 22 (2009).
 
 \bibitem{Schnyder2009} A. P. Schnyder, S. Ryu, A. Furusaki, and A. W. W. Ludwig, AIP Conf. Proc. {\bf 1134}, 10 (2009).
 
 \bibitem{Wen2014} X.-G. Wen, Phys. Rev. B {\bf 89}, 035147 (2014).
 
 \bibitem{Fu2007a} L. Fu, C. Kane, and E. Mele, Phys. Rev. Lett. {\bf 98}, 106803 (2007).
 
 \bibitem{Moore2007} J. Moore and L. Balents, Phys. Rev. B {\bf 75}, 121306(R) (2007).
 
 \bibitem{Roy2009} R. Roy, Phys. Rev. B {\bf 79}, 195322 (2009).
 
 \bibitem{Fu2011} L. Fu, Phys. Rev. Lett. {\bf 106}, 106802 (2011).
 
 \bibitem{Hsieh2012} T. H. Hsieh, H. Lin, J. Liu, W. Duan, A. Bansi, and L. Fu, Nature Commun. {\bf 3}, 982 (2012).
 
 \bibitem{Nomura2008} K. Nomura, S. Ryu, M. Koshino, C. Mudry, and A. Furusaki, Phys. Rev. Lett. {\bf 100}, 246806 (2008).
 
 \bibitem{Ringel2012} Z. Ringel, Y. E. Kraus, and A. Stern, Phys. Rev. B {\bf 86}, 045102 (2012).
 
 \bibitem{Mong2012} R. S. K. Mong, J. H. Bardarson, and J. E. Moore, Phys. Rev. Lett. {\bf 108}, 076804 (2012).
 
 \bibitem{Fu2012} L. Fu and C. L. Kane, Phys. Rev. Lett. {\bf 109}, 246605 (2012).
 
 \bibitem{Kobayashi2013} K. Kobayashi, T. Ohtsuki, and K.-I. Imura, Phys. Rev. Lett. {\bf 110}, 236803 (2013).
 
 \bibitem{Fulga2014} I. C. Fulga, B. van Heck, J. M. Edge, and A. R. Akhmerov, Phys. Rev. B {\bf 89}, 155424 (2014).
 
 \bibitem{Baireuther2014} P. Baireuther, J. M. Edge, I. C. Fulga, C. W. J. Beenakker, and J. Tworzyd{\l}o, Phys. Rev. B {\bf 89}, 035410 (2014).
 
 \bibitem{Sbierski2014} B. Sbierski and P. W. Brouwer, Phys. Rev. B {\bf 89}, 155311 (2014).
 
 \bibitem{Diez2015} M. Diez, D. I. Pikulin, I. C. Fulga, and J. Tworzyd{\l}o, New J. Phys. {\bf 17}, 043014 (2015).
 
 \bibitem{Morimoto2015} T. Morimoto, A. Furusaki, and C. Mudry, Phys. Rev. B {\bf 91}, 235111 (2015).
 
 \bibitem{Milsted2015} A. Milsted, L. Seabra, I. C. Fulga, C. W. J. Beenakker, and E. Cobanera, Phys. Rev. B {\bf 92}, 085139 (2015).
 
 \bibitem{Song2015} J. Song and E. Prodan, Phys. Rev. B {\bf 92}, 195119 (2015).
 
 \bibitem{Fulga2015} I. C. Fulga and M. Maksymenko, Phys. Rev. B {\bf 93}, 075405 (2016).
 
 \bibitem{Kennedy2015} R. Kennedy and C. Guggenheim, Phys. Rev. B {\bf 91}, 245148 (2015).
 
 \bibitem{Shiozaki2014} K. Shiozaki and M. Sato, Phys. Rev. B {\bf 90}, 165114 (2014).
 
 \bibitem{Baake2013} M. Baake and U. Grimm, \emph{Aperiodic Order Volume 1:
A Mathematical Invitation} (Cambridge University Press, 2013).

 \bibitem{BandresPhotonicQQ} M. A. Bandres, M. C. Rechtsman, and M. Segev, Phys. Rev. X {\bf 6}, 011016 (2016).
 
 \bibitem{Loring2015} T. A. Loring, Ann. Phys. {\bf 356}, 383 (2015).
 
 \bibitem{Fulga2011} I. C. Fulga, F. Hassler, A. R. Akhmerov, and C. W. J. Beenakker, Phys. Rev. B {\bf 83}, 155429 (2011).
 
 \bibitem{Fulga2012} I. C. Fulga, F. Hassler, and A. R. Akhmerov, Phys. Rev. B {\bf 85}, 165409 (2012).
 
 \bibitem{Yan2012} B. Yan, L. M{\"{u}}chler, and C. Felser, Phys. Rev. Lett. {\bf 109}, 116406 (2012).
 
 \bibitem{Rasche2013} B. Rasche, A. Isaeva, M. Ruck, S. Borisenko, V. Zabolotnyy, B. B{\"{u}}chner, K. Koepernik, C. Ortix, M. Richter, and J. van den Brink, Nature Mater. {\bf 12}, 422 (2013).
 
 \bibitem{Tang2014} P. Tang, B. Yan, W. Cao, S.-C. Wu, C. Felser, and W. Duan, Phys. Rev. B {\bf 89}, 041409 (2014).
 
 \bibitem{Yang2014} G. Yang, J. Liu, L. Fu, W. Duan, and C. Liu, Phys. Rev. B {\bf 89}, 085312 (2014).
 
 \bibitem{Pauly2015} C. Pauly, B. Rasche, K. Koepernik, M. Liebmann, M. Pratzer, M. Richter, J. Kellner, M. Eschbach, B. Kaufmann, L. Plucinski, C. M. Schneider, M. Ruck, J. van den Brink, and M. Morgenstern, Nature Phys. {\bf 11}, 338 (2015).
 
 \bibitem{Autes2015} G. Aut{\`e}s, A. Isaeva, L. Moreschini, J. C. Johannsen, A. Pisoni, R. Mori, W. Zhang, T. G. Filatova, A. N. Kuznetsov, L. Forr{\'o}, W. Van den Broek, Y. Kim, K. S. Kim, A. Lanzara, J. D. Denlinger, E. Rotenberg, A. Bostwick, M. Grioni, and O. V. Yazyev, Nature Mater. {\bf 15}, 154 (2015).
 
 \bibitem{Asahi2012} D. Asahi and N. Nagaosa, Phys. Rev. B {\bf 86}, 100504 (2012).
 
 \bibitem{Seroussi2014} I. Seroussi, E. Berg, and Y. Oreg, Phys. Rev. B {\bf 89}, 104523 (2014).
 
 \bibitem{Buehler2014} A. B\"{u}hler, N. Lang, C. Kraus, G. M\"{o}ller, S. Huber, and H. B\"{u}chler, Nature Commun. {\bf 5}, 4504 (2014).
 
 \bibitem{Diez2014} M. Diez, I. C. Fulga, D. I. Pikulin, J. Tworzyd{\l}o, and C. W. J. Beenakker, New J. Phys. {\bf 16}, 063049 (2014).
 
 \bibitem{Altland1997} A. Altland and M. R. Zirnbauer, Phys. Rev. B {\bf 55}, 1142 (1997).
 
 \bibitem{apptiling} The procedure used to construct the tiling is detailed
in the Supplemental Material, which contains Refs.\cite{Bhatia1997, Grunbaum1987, Beenker1982}.

 \bibitem{Bhatia1997} R. Bhatia, \emph{Matrix analysis} (Springer-Verlag, New York, 1997).
 
 \bibitem{Grunbaum1987} B. Gr\"{u}unbaum and G. C. Shephard, \emph{Tilings and patterns} (Freeman, 1987).
 
 \bibitem{Beenker1982} F. Beenker, \emph{Algebraic theory of non-periodic tilings of the plane by two simple building blocks: a square and a rhombus} (Eindhoven University of Technology, 1982).
 
 \bibitem{Groth2014} C. W. Groth, M. Wimmer, A. R. Akhmerov, and X. Waintal, New J. Phys. {\bf 16}, 063065 (2014).
 
 \bibitem{sourcecode} The source code of our numerical simulations is available online at URL....

 \bibitem{appsmatrix} In the Supplmental Material we describe the setup used to determine transport properties of the edge states.
 
 \bibitem{Tran2015} D.-T. Tran, A. Dauphin, N. Goldman, and P. Gaspard, Phys. Rev. B {\bf 91}, 085125 (2015).
 
 \bibitem{Kitaev2001} A. Y. Kitaev, Physics-Uspekhi {\bf 44}, 131 (2001).
 
 \bibitem{Brouwer2000} P. W. Brouwer, A. Furusaki, I. A. Gruzberg, and C. Mudry, Phys. Rev. Lett. {\bf 85}, 1064 (2000).
 
 \bibitem{Motrunich2001} O. Motrunich, K. Damle, and D. A. Huse, Phys. Rev. B {\bf 63}, 224204 (2001).
 
 \bibitem{Brouwer2003} P. W. Brouwer, A. Furusaki, and C. Mudry, Phys. Rev. B {\bf 67}, 014530 (2003).
 
 \bibitem{Gruzberg2005} I. A. Gruzberg, N. Read, and S. Vishveshwara, Phys. Rev. B {\bf 71}, 245124 (2005).
 
 \bibitem{Essin2007} A. M. Essin and J. E. Moore, Phys. Rev. B {\bf 76}, 165307 (2007).
 
 \bibitem{Kitaevlecture} A. Kitaev, “Classification of topological insulators and superconductors,” Lecture given at the IPMU focus week “Condensed Matter Physics Meets High Energy Physics,” University of Tokyo, 2010, pp. 8–12.
 
 \bibitem{Akhmerov2011} A. R. Akhmerov, J. P. Dahlhaus, F. Hassler, M. Wimmer, and C. W. J. Beenakker, Phys. Rev. Lett. {\bf 106}, 057001 (2011).
 
 \bibitem{nopbchere} Here, the scattering matrix weak invariant can be obtained without periodic boundary conditions, since no edge modes connect the top and bottom leads.
 
 \bibitem{Kellendonk2015} J. Kellendonk, arXiv:1509.06271 (2015).
\end{thebibliography}
\end{document}


\title{Supplemental Material to: ``Aperiodic weak topological superconductors''}

\author{I. C. Fulga}
\affiliation{Department of Condensed Matter Physics, Weizmann Institute of Science, Rehovot 76100, Israel}
\author{D. I. Pikulin}
\affiliation{Department of Physics and Astronomy and Quantum Matter Institute, University of British Columbia, Vancouver, BC, Canada V6T 1Z1}
\author{T. A. Loring}
\affiliation{Department of Mathematics and Statistics, University of New Mexico, Albuquerque, NM 87131, USA}

\maketitle
\appendix

\section{Proofs}
\label{app:proofs}

Here we provide a more detailed analysis of the pseudospectrum invariants used in the main text, and prove the existence and stability of localized zero energy modes in the weak phase. Consider a generic finite system in symmetry class D, with geometry a square of length $L$ centered at the origin. Let $H$ be the Hamiltonian of the system, which can be brought to an imaginary form in class D, $H=-H^*$, for instance by a basis change of the form (6) in the main text. We denote by $X$ and $Y$ the position operators associated to the system, which are real, commute with each other, and obey $-L/2 \leq X, Y \leq L/2$. Additionally, $X$ and $Y$ almost commute with $H$, roughly meaning that $\|[H,X]\|/\|X\|$ and $\|[H,Y]\|/\|Y\|$ are less than the gap in the bulk spectrum, where $\|.\|$ denotes the operator norm or 2-norm. We consider three different pseudo-spectra. The first uses

\begin{equation} \label{eq:ps_b_st}
 B_{\bl}^{\rm st} = B(X-\lambda_1 I, Y-\lambda_2 I, H - \lambda_3 I)
\end{equation}
where
\begin{equation} \label{eq:def_B}
 B(A_1, A_2, A_3) = \begin{pmatrix}
		      A_1 & A_2 - i A_3 \\
		      A_2 + i A_3 & -A_1
                    \end{pmatrix}.
\end{equation}

In Eq.~\eqref{eq:ps_b_st}, the real valued parameters $\lambda_1$ and $\lambda_2$ are shifts in the $x$ and $y$ coordinates, while $\lambda_3$ corresponds to a shift in energy. Using this expression, we define the strong gap function
\begin{equation} \label{eq:def_gap_st}
 {\rm gap}_\bl^{\rm st} = \sigma_{\rm min} \left( B_\bl^{\rm st} (X, Y, H) \right),
\end{equation}
where $\sigma_{\rm min}$ stands for the smallest singular value and $\bl =(\lambda_1,\lambda_2,\lambda_3)$. The function gap$_\bl^{\rm st}$ can be used to construct the pseudospectrum. For a given $\varepsilon \geq 0$ ($\geq$ instead of $>$ is to make the
$\varepsilon$-pseudospectrum a compact	set), the $\varepsilon$-pseudospectrum
of $(X, Y, H)$, which we call in this appendix the {\em strong
$\varepsilon$-pseudospectrum}, is the set
\begin{equation} \label{eq:def_st_ps}
 \Lambda_\varepsilon = \left\{ \bl\in\mathbb{R}^3 \,|\, {\rm gap}_\bl^{\rm st} \leq \varepsilon \right\}.
\end{equation}

As discussed in Ref.~\cite{Loring2015}, the strong  pseudospectrum can be used to construct a topological invariant leading to gapless modes on all edges of the system. As long as $\bl\notin\Lambda_{0} (X, Y, H)$, the invariant is the integer
\begin{equation} \label{eq:def_st_inv}
 {\rm Ind}_\bl^{\rm st} = \frac{1}{2}\, {\rm Sig} \left( B_\bl^{\rm st} (X, Y, H) \right)
\end{equation}
where ${\rm Sig}$ stands for matrix signature. In practice, we require $\bl\notin\Lambda_{\epsilon} (X, Y, H)$ for some small $\epsilon$ to make the computation of the signature numerically stable. The expression \eqref{eq:def_st_inv} is a generalized version of the invariant (7) used in the main text, as it also takes into account shifts in position or energy, which are parametrized by the vector $\bl = (\lambda_1, \lambda_2, \lambda_3)$.

\begin{figure}
\includegraphics[width=0.7\columnwidth]{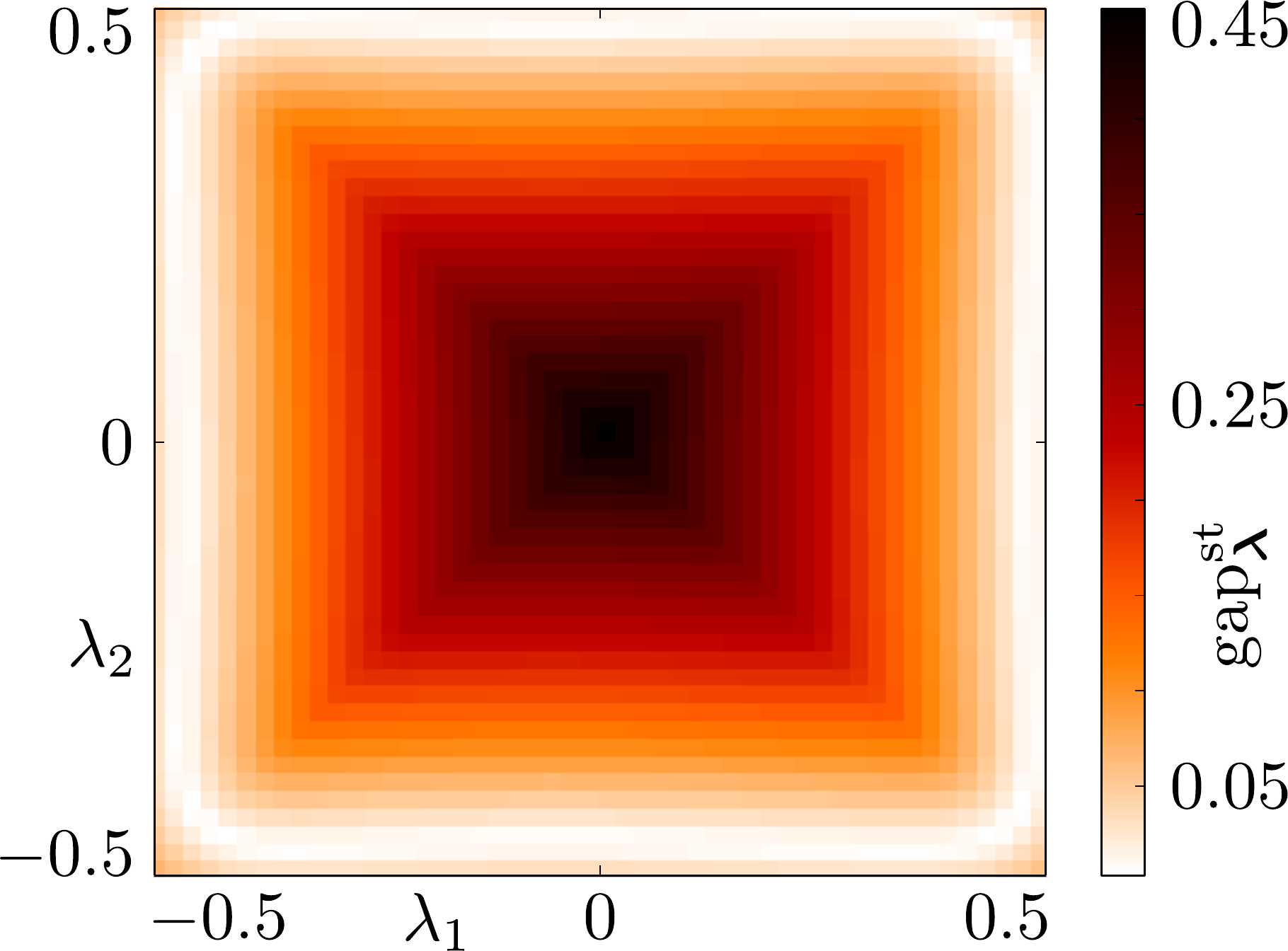}
\caption{Plot of $\mathrm{gap}_{\bl}^{\mathrm{st}}$ as a function of $\bl=(\lambda_{1},\lambda_{2},0)$ for the TI phase of Fig.~1 in the main text. It takes large values in the bulk of the system and decreases towards the boundary, reaching a minimum computed value of approximately $3\times10^{-3}$. \label{fig:Plot-strong-gap}}
\end{figure}

Figure \ref{fig:Plot-strong-gap} shows a plot of $\mathrm{gap}_{\bl}^{\mathrm{st}}$ for the Hamiltonian $H_{QC}$ with the parameters of of the strong phase (Fig.~1 in the main text). The gap function is computed at the Fermi level and as a function of position, that is to say using $\bl = (\lambda_1, \lambda_2, 0)$.
We computed the index at the origin, and it cannot change where the gap remains open, so in index is $-1$ within the square. It is undefined, or defined and meaningless, near the edges.

The two weak invariants are created by ignoring one position observable and using the one-dimensional expressions introduced in Ref.~\cite{Loring2015}. We only need one of these, the one that ignores the $x$ position. Define
\begin{equation} \label{eq:def_b_y}
 B_{\bl}^{y}(X,Y,H)=B(0,Y-\lambda_{1}I,H-\lambda_{2}I),
\end{equation}
which is Hermitian, as we always assume the $\lambda_{j}$ are real, and real symmetric when $\lambda_{2}=0$. We consider the function 
\begin{equation} \label{eq:def_gap_y}
 \mathrm{gap}_{\bl}^{y}=\sigma_{\mathrm{min}}\left(B_{\bl}^{y}(X,Y,H)\right).
\end{equation}
The {\em weak  $\epsilon$-pseudospectrum} of $(X,Y,H)$ is the set 
\begin{equation} \label{eq:def_weak_ps}
 \Lambda_{\epsilon}^{y}(X,Y,H)=\left\{ \bl\in\mathbb{R}^{2}\left|\mathrm{gap}_{\bl}^{y}\leq\epsilon\right.\right\} .
\end{equation}

\begin{figure}
\includegraphics[width=\columnwidth]{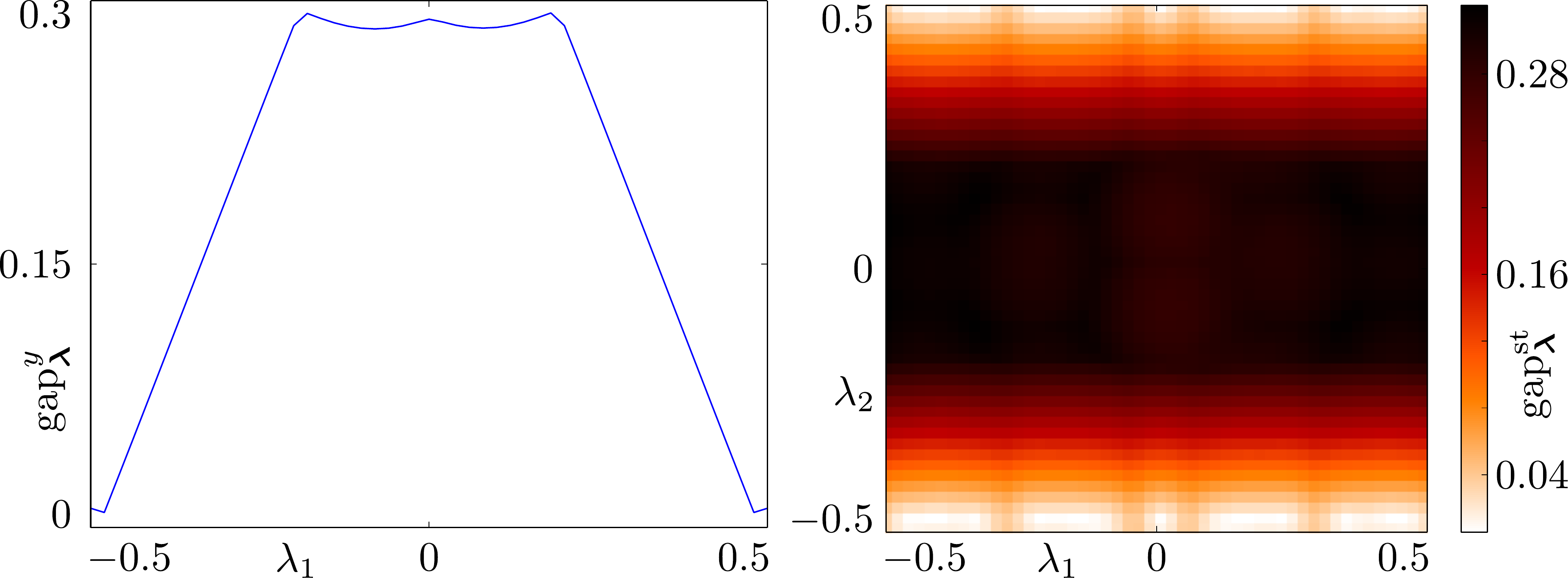}
\caption{Left: Plot of $\mathrm{gap}_{\bl}^{y}$ as a function of $\bl=(\lambda_{1},0)$ for the WTI phase (Fig.~2 in the main text), in the absence of disorder. Right: Corresponding plot of $\mathrm{gap}_{\bl}^{\rm st}$ as a function of $\bl=(\lambda_{1},\lambda_2, 0)$. Both $\mathrm{gap}_{\bl}^{y}$ and $\mathrm{gap}_{\bl}^{\rm st}$ take large values in the bulk and decrease towards the edge, reaching a minimum computed value of approximately $86\times10^{-4}$ for $\mathrm{gap}_{\bl}^{\rm st}$ and $56\times10^{-4}$ for $\mathrm{gap}_{\bl}^{y}$ at $y \simeq -0.48 $ and $y \simeq 0.48$. \label{fig:Plot-y-gap}}
\end{figure}

In Fig.~\ref{fig:Plot-y-gap} we plot gap$_\bl^{\rm st}$ and $\mathrm{gap}_{\bl}^{y}$ for Hamiltonian parameters of the weak phase (Fig.~2 in the main text). The gap functions are computed at zero energy and for different $x$ and $y$-locations.
A non-trivial index at a large value of $\mathrm{gap}_{\bl}^{y}$ forces the $\mathrm{gap}_{\bl}^{y}$ to hit zero at some $y$-value to the left and the right. Furthermore, when $\mathrm{gap}_{\bl}^{y}$ hits zero there must be a corresponding state localized in energy and $y$-position. This means the approximate zero modes are topologically protected against some forms of disorder. We make these statements precise in the next two Theorems:

\begin{thm}\label{thm:gap_closing_and_deformation}
Assume $(X,Y,H)$ are as defined in this section. If, for some $ $$\boldsymbol{\mu}=(\mu_{1},0)$
with $\mu_{1}>0$, we have $\mathrm{gap}_{\boldsymbol{\mu}}^{y}=C$
and $\mathrm{Ind}_{\boldsymbol{\mu}}^{y}(X,Y,H)=-1$ , then for
any imaginary Hermitian matrix $K$ with $\|H-K\|\leq C$ there will
exist $\lambda_{1}>\mu_{1}$ so that $\mathrm{gap}_{(\lambda_{1},0)}^{y}=0$. 

A symmetric statement holds when $\mu_{1}<0$.
\end{thm}

\begin{proof}
From $\|H-K\|\leq C$ we conclude that 
\[
\|B_{\boldsymbol{\mu}}^{y}(X,Y,H)-B_{\boldsymbol{\mu}}^{y}(X,Y,K)\|<C
\]
and so, by Weyl's estimate on spectral variation \cite{Bhatia1997}, the gap function defined using $B_{\boldsymbol{\mu}}^{y}(X,Y,K)$ is nonzero. This gap remains open if we interpolate linearly between $H$ and $K$, so the weak index cannot become positive along this path: 
$
\mathrm{Ind}_{\boldsymbol{\mu}}^{y}(X,Y,K)=-1
$.
By Theorem 7.5 in Ref.~\onlinecite{Loring2015} there is $\lambda_{1}>\mu_{1}$
so that $\mathrm{gap}_{(\lambda_{1},0)}^{y}=0$. 
\end{proof}

\begin{thm}\label{thm:localized_zero_modes}
Assume $(X,Y,H)$ are as defined in this section. Suppose \textup{$\epsilon=\mathrm{gap}_{\bl}^{y}$ for some }$\bl=(\lambda_{1},0)$. Then there is a real unit vector $\mathbf{v}$ so that 
\[
\left\Vert Y\mathbf{v}-\lambda_{1}\mathbf{v}\right\Vert ,\left\Vert H\mathbf{v}\right\Vert \leq\sqrt{\epsilon^{2}+\left\Vert \left[Y,H\right]\right\Vert }.
\]
\end{thm}

\begin{proof}
By Lemma 1.2 in \cite{Loring2015} there is a unit vector $\mathbf{v}$
so that 
\[
\left\Vert Y\mathbf{v}-\lambda_{1}\mathbf{v}\right\Vert ,\left\Vert H\mathbf{v}\right\Vert \leq\sqrt{\epsilon^{2}+\left\Vert \left[Y,H\right]\right\Vert }.
\]
In fact, the proof of Lemma 1.2 in Ref.~\onlinecite{Loring2015} finds such a $\mathbf{v}$ by selecting $\mathbf{v} = \mathbf{v}_j$, the larger of the top or bottom half of any eigenvector 
\[
\mathbf{w}=\left[\begin{array}{c}
\mathbf{v}_{1}\\
\mathbf{v}_{2}
\end{array}\right]
\]
of $B_{\bl}^{y}(X,Y,H)$ for an eigenvalue $\alpha$ that is closest to $0$, and rescaling \textbf{v} to be a unit vector.  We know this eigenvector can be selected to be real, so we can assume $\mathbf{v}$ is also real.
\end{proof}

To translate results about approximate eigenstates into more standard physics terminology, we offer the following Lemmas. 

\begin{lem}
If $\|X\mathbf{v}-\lambda\mathbf{v}\|\leq\delta$ for some Hermitian
matrix $X$ and unit vector $\mathbf{v}$ then
\[
\left\langle \mathbf{v}|X^{2}|\mathbf{v}\right\rangle -\left\langle \mathbf{v}|X|\mathbf{v}\right\rangle ^{2}
\leq 2\delta^{2}
\]
 and 
\[
\lambda-\delta \leq 
\left\langle \mathbf{v}|X |\mathbf{v}\right\rangle 
\leq\lambda+\delta.
\]
\end{lem}

\begin{proof}
We compute
\[
\left|\strut\left\langle \mathbf{v}| X |\mathbf{v}\right\rangle - \lambda \  \strut\right|
=
\left|\strut \left\langle X \mathbf{v}-\lambda\mathbf{v} | \mathbf{v}\right\rangle \strut \right|
\]
and (by the law of cosines) 
\begin{align*}
\left|\strut\left\langle \mathbf{v}|X^{2}|\mathbf{v}\right\rangle -\lambda^{2} \ \strut\right|
& =
\left|\strut\|X\mathbf{v}\|^{2}-\|\lambda\mathbf{v}\|^{2}\strut\right|
\\
& \leq
\|\strut X \mathbf{v} -\lambda\mathbf{v}\strut\|^{2}.
\end{align*}
To prove the first inequality we use the shift invariance of variance to reduce to the case $\lambda=0$, where
\[
\left\langle \mathbf{v}|X^2|\mathbf{v}\right\rangle - \left\langle\mathbf{v}|X|\mathbf{v}\right\rangle ^{2}
\leq
\left\langle \mathbf{v}|X^2|\mathbf{v}\right\rangle +
\left\langle \mathbf{v}|X|\mathbf{v}\right\rangle ^{2}
\leq
2\delta^{2}.
\]
\end{proof}

\begin{lem}
If $H$ is Hermitian and imaginary, and if $\mathbf{v}$ is a real unit vector, then 
$\left\langle \mathbf{v}| H |\mathbf{v}\right\rangle = 0$.
\end{lem}

\begin{proof}
We simply conjugate $\left\langle \mathbf{v} | H |\mathbf{v}\right\rangle$ and compute:
\[
\left\langle \mathbf{v}|H|\mathbf{v}\right\rangle 
=
\overline{\left\langle \mathbf{v}|H|\mathbf{v}\right\rangle }
=
\left\langle \overline{\mathbf{v}}|\overline{H}|\overline{\mathbf{v}}\right\rangle 
=
-\left\langle \mathbf{v}|H|\mathbf{v}\right\rangle .
\]
\end{proof}

Using the above Lemmas and Theorems, we summarize the stability of the weak topological phase as the following Corollary. It is stated for on-site disorder, correlated or not.  It can be modified to give slightly weaker estimates for more general forms of local disorder.  A symmetric result holds for negative $y$-values.

\begin{cor} \label{thm:WTI_makes_edge_modes}
Let $H_{{\rm QC}}$ denote the Hamiltonian constructed with the parameters used in Fig.~2 of the main text.  Fig.~\ref{fig:Plot-y-gap} illustrates $\mathrm{gap}_{\bl}^{y}$ for this Hamiltonian. Let $\delta=\left\Vert \left[H_{{\rm QC}},Y\right]\right\Vert \simeq 0.0757$, as obtained using our choice of parameters. 
Suppose $H_{{\rm dis}}$ is Hermitian with the same particle-hole symmetry as $H_{{\rm QC}}$, and $H_{{\rm dis}}-H_{{\rm QC}}$ consists only of on-site disorder.  
If disorder is not too large, meaning
\[
\left\Vert H_{{\rm dis}}-H_{{\rm QC}}\right\Vert \leq\mathrm{gap}_{(y_{0}+\delta,0)}^{y}
\]
for some position to the right of the origin, $0 \leq y_{0} \leq 0.48-\sqrt{\delta}$, then there exists a normalized, particle-hole symmetric state $\psi$  with the following properties:
\begin{itemize}
 \item The state $\psi$ is at zero energy, $\langle \psi|H_{\rm dis}|\psi\rangle=0$, with a variance $\langle \psi |H_{\rm dis}^2| \psi \rangle - \langle \psi | H_{\rm dis} | \psi \rangle^2\equiv\Delta_{\psi}^{2}H_{{\rm dis}}\leq2\delta^{2}$
 \item The state $\psi$ is localized in position to the right of the origin point, $y_{0}\leq\langle\psi|Y|\psi\rangle$ and $\Delta_{\psi}^{2}Y\leq2\delta^{2}$.
\end{itemize}
\end{cor}

\section{Generating the Ammann-Beenker tiling}

\begin{figure*}[t]
\includegraphics[width=0.7\linewidth]{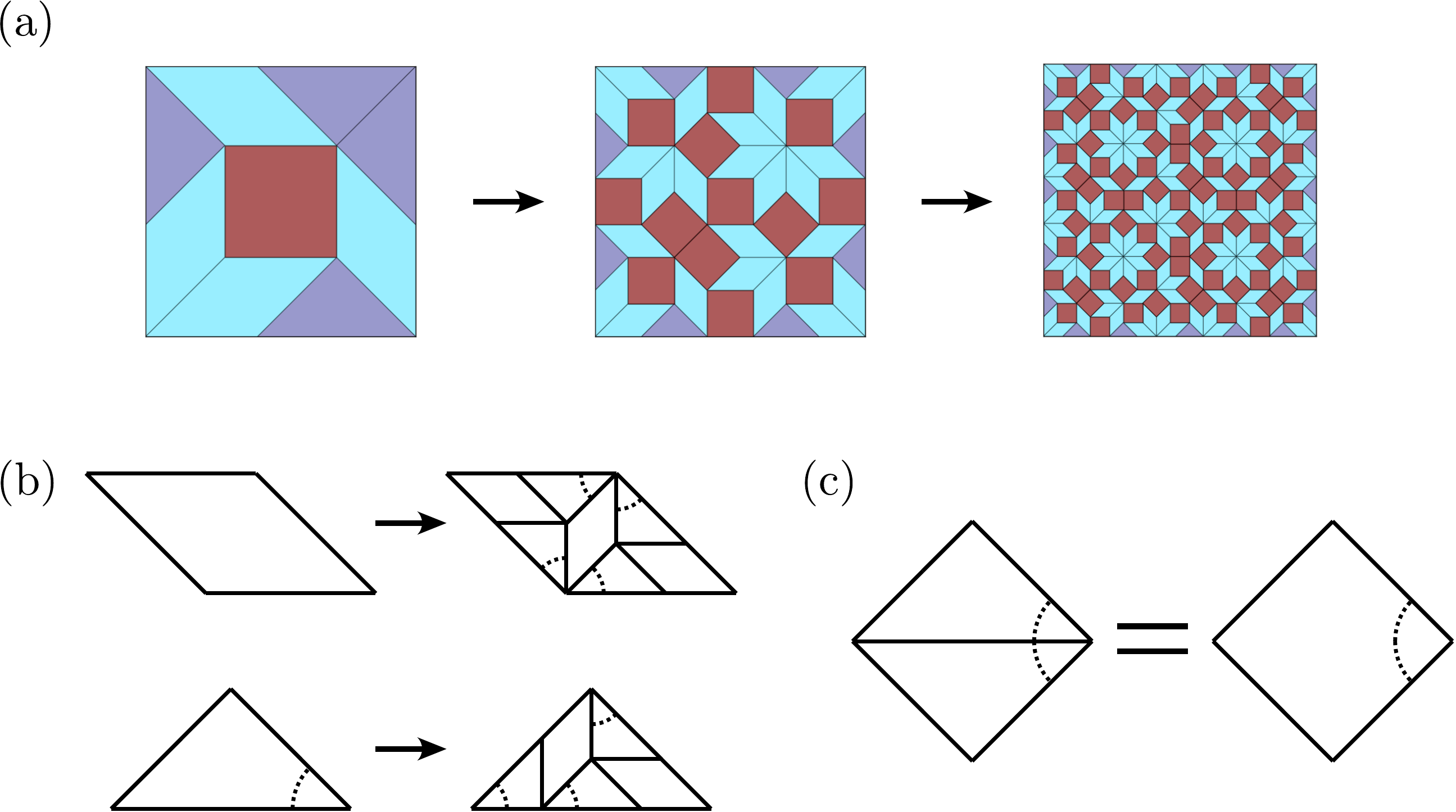}
\caption{Obtaining the Ammann-Beenker tiling through subdivision. (a) Example of the subdivision process. (b) Elementary step of the subdivision of the rhombus and the (decorated) triangle. (c) Equivalence between two triangles and a square tile, which is applied in the beginning and the end of the subdivision process. \label{fig:subdivisions}}
\end{figure*}

Ammann was the first to suggest a tiling of the plane respecting perfect 8-fold rotational symmetry \cite{Grunbaum1987}. It was later put forward by Beenker that this tiling can be obtained by subdivision of the rhombus with angles $\pi/4$ and $3\pi/4$, and the square \cite{Beenker1982}. We use a modified version of this algorithm to generate the tiling shown in the main text. An example of a three-step subdivision procedure is shown in Fig.~\ref{fig:subdivisions}a. The initial step of the algorithm consists of converting a square shape into two decorated triangles, as shown in Fig.~\ref{fig:subdivisions}c. Afterwards, we iteratively apply the subdivision rules shown in Fig.~\ref{fig:subdivisions}b, each time obtaining a quasicrystal patch with a larger number of sites. Finally, at the end of the procedure we convert all pairs of decorated triangles back into squares, by applying the equivalence of Fig.~\ref{fig:subdivisions}c. The resulting quasicrystal patches are shown in the main text figures. We have used three subdivisions to generate the tiling shown in Fig.~1 of the main text, and four subdivisions for the tiling in Fig.~2.

\section{Transport setup}

Here we give details on the transport setup used to determine the conducting properties of Majorana edge states, both in the strong and in the weak topological phase.

\begin{figure}[htb]
\includegraphics[width=0.7\linewidth]{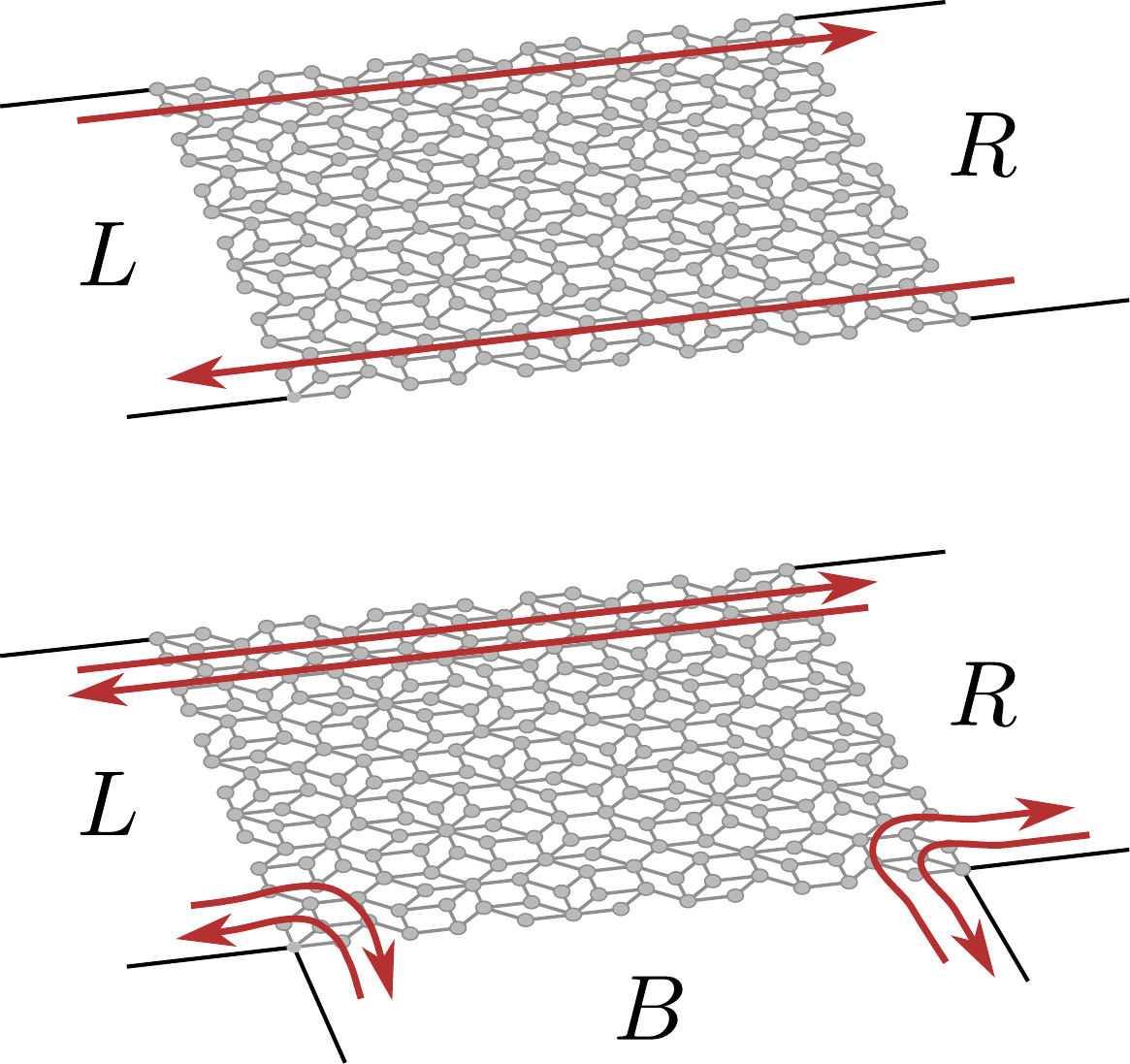}
\caption{Setup used to compute heat transport through the system. Top panel: In the strong topological phase, we attach leads (labeled $L$ and $R$) to the left- and right-most sites of the tiling. Bottom panel: in the weak phase, to determine the heat conductance of only the top edge states, we attach a third lead to the bottom-most sites (labeled $B$).  \label{fig:transport}}
\end{figure}

In the strong topological phase, the aperiodic system $H_{\rm QC}$ has an insulating bulk and chiral Majorana edge states which conduct heat. We attach two leads, to the left- and right-most sites of the tiling, as shown in the top panel of Fig.~\ref{fig:transport}, leading to a scattering matrix of the form (5) in the main text. The only contribution to thermal conductance comes from the chiral edge states, leading to a quantized thermal conductance equal to the Chern number. For the parameters used in Fig.~1 of the main text, we find $G/G_0=1$.

In the weak topological phase, we study the properties of a single edge and compare its conductance distribution with the prediction (8) in the main text. To this end, we attach a third lead to the bottom-most sites of the tiling, as shown in the bottom panel of Fig.~\ref{fig:transport}. In this setup, transport from the left to the right lead, labeled $L$ and $R$, only occurs through the top pair of counter-propagating edge states.